\documentclass[aps,prl,reprint,showpacs,superscriptaddress,floatfix]{revtex4-1}
\usepackage{graphicx}
\usepackage{color}
\usepackage{amsmath}
\usepackage[hidelinks]{hyperref}

\begin{document}

\title{Microscopic evidence for strong periodic lattice distortion in two-dimensional charge-density wave systems}

\author{Jixia~Dai}
\email[]{daij@colorado.edu}
\affiliation{Department of Physics, University of Colorado at Boulder, Boulder, CO 80309, USA}

\author{Eduardo~Calleja}
\affiliation{Department of Physics, University of Colorado at Boulder, Boulder, CO 80309, USA}

\author{Jacob~Alldredge}
\affiliation{Department of Physics, University of Colorado at Boulder, Boulder, CO 80309, USA}

\author{Xiangde~Zhu}
\affiliation{High Magnetic Field Laboratory, CAS, Hefei 230031, China}

\author{Lijun~Li}
\affiliation{Key Laboratory of Materials Physics, Institute of Solid State Physics, CAS, Hefei 230031, China}

\author{Wenjian~Lu}
\affiliation{Key Laboratory of Materials Physics, Institute of Solid State Physics, CAS, Hefei 230031, China}

\author{Yuping~Sun}
\affiliation{Key Laboratory of Materials Physics, Institute of Solid State Physics, CAS, Hefei 230031, China}
\affiliation{High Magnetic Field Laboratory, CAS, Hefei 230031, China}

\author{Thomas~Wolf}
\affiliation{Karlsruher Institut f\"{u}r Technologie, Institut f\"{u}r Festk\"{o}rperphysik, P.O.B. 3640, D-76021 Karlsruhe, Germany}

\author{Helmuth~Berger}
\affiliation{Institute of Physics of Complex Matter, Ecole Polytechnique Federal de Lausanne, CH-1015 Lausanne, Switzerland}

\author{Kyle~McElroy}
\email[]{Kyle.McElroy@colorado.edu}
\affiliation{Department of Physics, University of Colorado at Boulder, Boulder, CO 80309, USA}

\begin{abstract}
In the quasi-2D electron systems of the layered transition metal dichalcogenides (TMD) there is still a controversy about the nature of the transitions to charge-density wave (CDW) phases, i.e. whether they are described by a Peierls-type mechanism or by a lattice-driven model.
By performing scanning tunneling microscopy (STM) experiments on the canonical TMD-CDW systems, we have imaged the electronic modulation and the lattice distortion separately in 2H-TaS$_\mathrm{2}$, TaSe$_\mathrm{2}$, and NbSe$_\mathrm{2}$. Across the three materials, we found dominant lattice contributions instead of the electronic modulation expected from Peierls transitions, in contrast to the CDW states that show the hallmark of contrast inversion between filled and empty states.
Our results imply that the periodic lattice distortion (PLD) plays a vital role in the formation of CDW phases in the TMDs and illustrate the importance of taking into account the more complicated lattice degree of freedom when studying correlated electron systems.

\end{abstract}

\pacs{71.45.Lr, 71.27.+a, 74.55.+v}
\maketitle

The transition metal dichalcogenides (TMD) have shown intriguing phenomena like the spin-valley physics \cite{Xiao2012,Wu2013}, superconductivity induced/enhanced by tuning various parameters \cite{Ye2012a,Taniguchi2012,Sipos2008,Morosan2006,Wagner2008,Li2010,Friend1987}, and charge-density waves (CDW) that coexist and compete with superconductivity (e.g. 2H-$\mathrm{NbSe_2}$, $\mathrm{TaSe_2}$, 1T-$\mathrm{TaS_2}$, and $\mathrm{TiSe_2}$) \cite{Sipos2008,Morosan2006,Wagner2008,Li2010,Friend1987}. 
Such competition between CDW and superconductivity is closely tied to the complicated interactions between the internal degrees of freedom, including charge, lattice and orbital.
Differentiating which interactions are key requires knowing which of them are responsible for the different phases. The degree to which an electron-driven mechanism is the cause of CDWs in the trigonal prismatic structured (2H) TMDs is still under debate.
Existing studies have proposed a variety of different mechanisms, including the Fermi surface nesting \cite{Borisenko2008,Borisenko2009}, saddle band driven susceptibility divergence \cite{Rice1975}, f-wave gapping and marginal Fermi liquid \cite{CastroNeto2001}, etc.

These materials do show some evidence of the electronic nature of the CDW phases including the existence of incommensurate CDW phases \cite{Moncton1977,Scholz1982}, since $2k_F$ is generally not expected to be a rational fraction of the lattice reciprocal vectors.
Furthermore, the electronic origin is also supported by photoemission experiments which measured the Fermi surface of $\mathrm{NbSe_2}$ and $\mathrm{TaSe_2}$ and approximated the electronic susceptibility with an autocorrelation. This analysis showed peaks at wave vectors corresponding to those of the CDW \cite{Borisenko2008,Borisenko2009}.
However, inelastic X-ray scattering experiment in $\mathrm{NbSe_2}$ \cite{Weber2011} reveals that the lattice dynamics exhibit unconventional behavior and hence may be dominating the transition.
Moreover, recent LDA calculations have shown the difficulties with Fermi surface nesting and have suggested that periodic lattice distortion (PLD), instead of eMod, is the essential ingredient \cite{Johannes2006,Johannes2008}.
In a recent real space study, Soumyanarayanan \emph{et al.} \cite{Soumyanarayanan2013} have shown that there is a close relationship between CDW formation and the local strain related to the surface layer buckling in $\mathrm{NbSe_2}$.
Rossnagel \emph{et al.} have, in addition, argued that the spin-orbit interaction is also important for a 5d-electron system like $\mathrm{TaSe_2}$ \cite{Rossnagel2007}.
Because of this debate, it is necessary to systematically study the CDWs in the TMDs with the simultaneous measurement of any intrinsic lattice distortion while spatially resolving the electronic component across this family of compounds.

The CDW instability in low dimensional systems was initially explained by the Peierls mechanism \cite{Peierls1955,Gruner1994,Johannes2008} in which a divergence in electronic response function at a particular wave vector ($2k_\mathrm{F}$ for 1D) results in a periodic charge redistribution in real space.
That is, a low dimensional electronic system can lower its energy by opening an energy gap at the Fermi level, piling up electrons periodically and reducing its lattice transitional symmetry.
As a result of the periodic electron modulation (eMod) and the electron-phonon interaction, the lattice itself undergoes a similar modulation, which causes an increase in the lattice elastic energy \cite{Gruner1994}.
As a consequence, with this mechanism the lattice distortion is expected to be minimized acting like a competing effect \cite{Johannes2008}. The eMod manifests in the change of the local density of states (LDOS) shows the periodic increase in density of electron with holes modulating with the same period but out of phase with the electrons. This can be shown by calculating the LDOS in real-space using the Peierls model \footnote{See Supplemental Material for the detailed LDOS derivation}.
Another way of visualizing this alternating intensity is by using the quantum mechanical sum rule, where the summation of the number of filled and empty states on every lattice site is a constant.
This means that wherever there are excess electron states, there must be fewer hole states on the same site.
The LDOS calculation we have performed also shows this conservation of total states.

To investigate the nature of the CDW in these materials we use Scanning Tunneling Microscopy (STM) in which the tunneling current is proportional to the sample's integrated density of states and a matrix element with exponential dependence on the tip-sample distance, $I\sim e^{-2\kappa s}\!\int_0^{eV}\! LDOS(\epsilon)d\epsilon$ \cite{Chen1993}.
In the constant current imaging mode, the tunneling current is maintained at a fixed value with the tip-sample distance being adjusted and recorded.
For a CDW system that only contains eMod and no lattice distortions, the topographic images of the filled and empty states will be complementary with each other due to the nature of LDOS variation and uniformity of atomic heights.
This phenomenon has been well demonstrated by STM experiments on the quasi-1D $\mathrm{NbSe_3}$ \cite{Brun2009}, the quasi-2D system of $\mathrm{K_{0.9}MoO_{3}}$ \cite{Mallet1999}, and the quasi-2D $\mathrm{CaC_6}$ with 1D modulation \cite{Rahnejat2011}.
In each of these cases the modulations for filled and empty states, with the CDW periodicity, are of opposite phases in the topography (contrast inversion) as predicted by the Peierls model.

\begin{figure}[t]
  \includegraphics[width=3.25in]{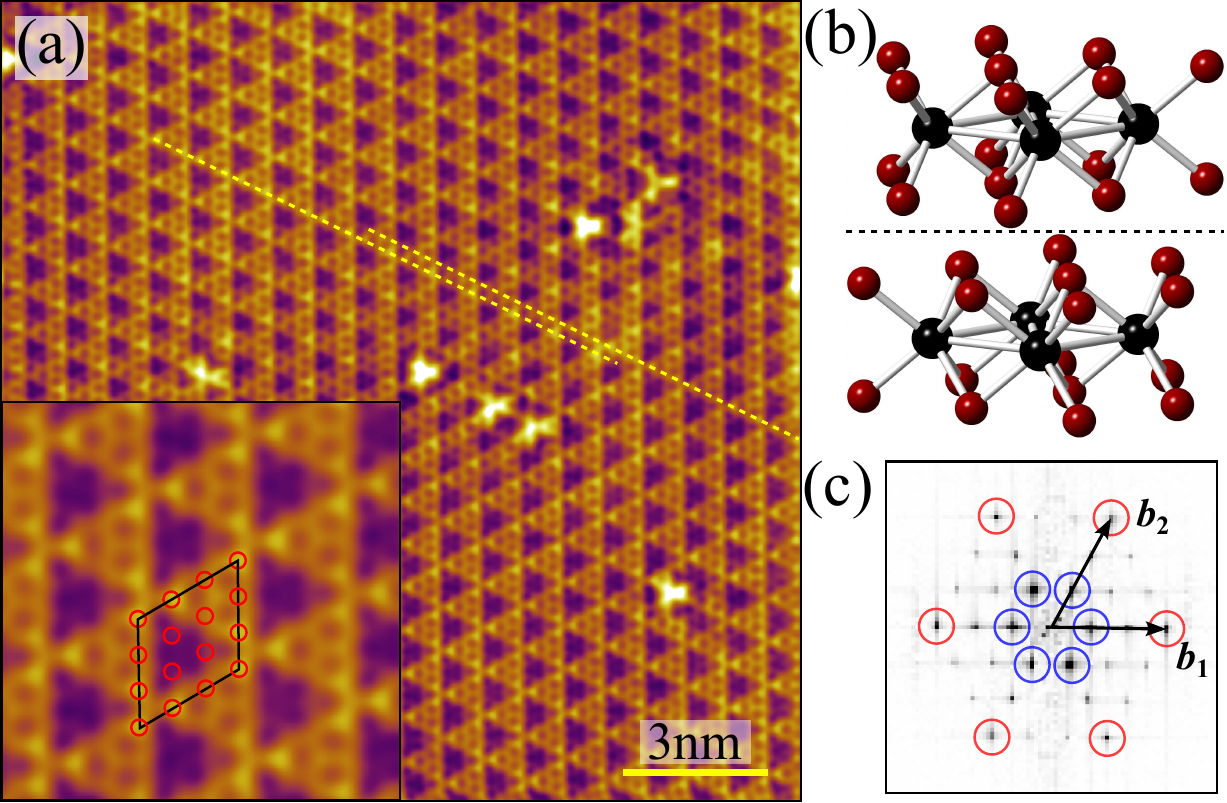}\\
  \caption{\label{fig1}(Color online) (a) Atomically resolved topographic image of 2H-$\mathrm{TaS_2}$ with CDW modulation (3 $\times$ 3 superstructure). This image was obtained at 60~K and with setup condition being $-$100~mV and 200~pA. Inset: zoom in to show the atomic structure of the 3 $\times$ 3 unit cell with red circles indicating the sulfur atoms. (b) Crystal structure of trigonal prismatic (2H) TMD. The red/black atoms are chalcogen/metal atoms, and the dashed line is where it cleaves. (c) Fourier transform of (a). Blue (red) circles indicate the primary peaks of CDW (atomic) modulation, and $\boldsymbol{b_1}$ and $\boldsymbol{b_2}$ are the atomic lattice reciprocal vectors.}
\end{figure}

\begin{figure}[b]
  \includegraphics[width=3.25in]{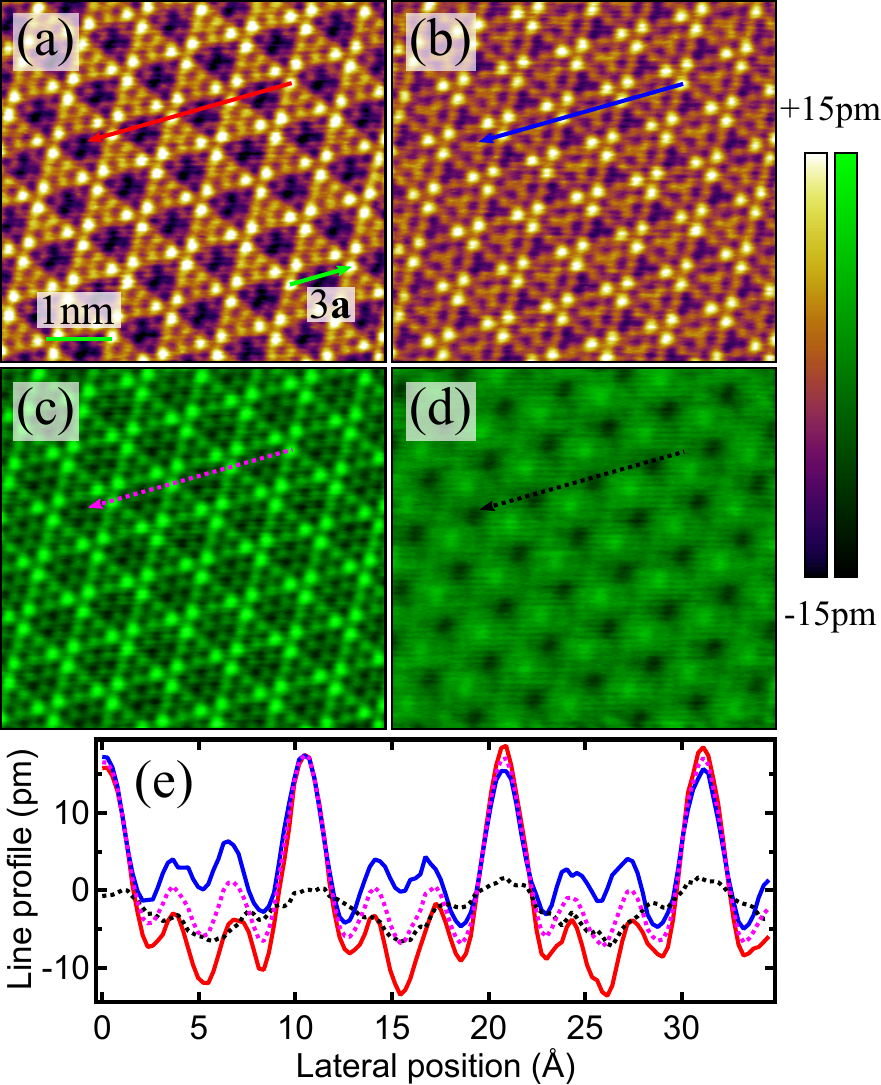}\\
  \caption{\label{fig2}(Color online) Topographic images showing symmetrization and antisymmetrization on 2H-$\mathrm{TaS_2}$ at 52~K. (a) and (b): Topographic images with $-/+$50~mV and 100~pA. These two images were individually corrected for drift and aligned with sub-atomic precision. (c) and (d): The symmetrized ($S = (a+b)/2$) and anti-symmetrized ($AS = (a-b)/2$) images of (a) and (b). The contrast in (d) is much smaller compared to (c). (e): The line profiles in (a-d) indicated by colors.}
\end{figure}

\begin{figure*}[t]
  \includegraphics[width=6.5in]{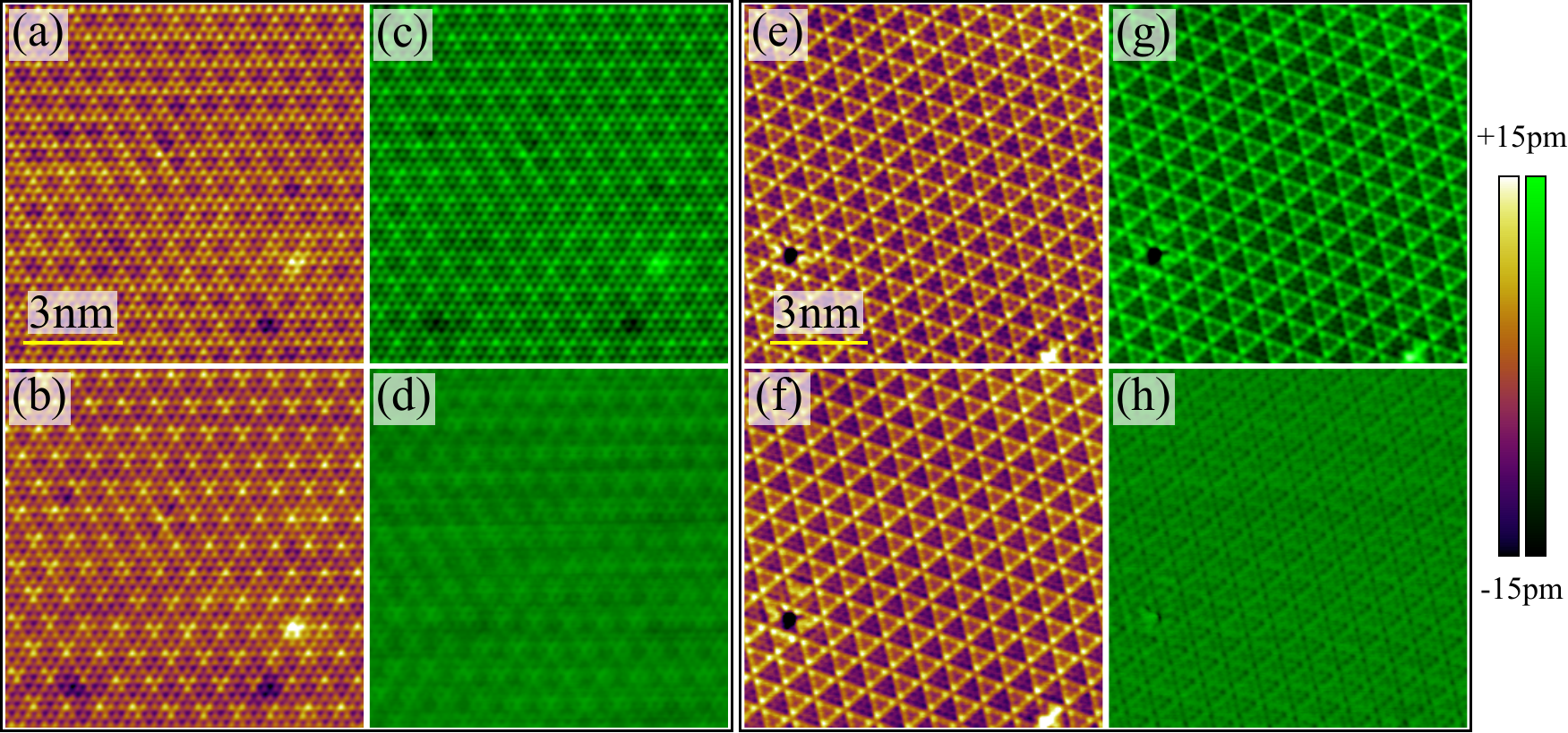}\\
  \caption{\label{fig3}(Color online) Topographic images of $\mathrm{NbSe_2}$ and $\mathrm{TaSe_2}$ at 5~K. (a/b): Topographic images of $\mathrm{NbSe_2}$ with $-/+$100~mV and 100~pA. (e/f): topographic images of $\mathrm{TaSe_2}$ with $-/+$50~mV and 40/30~pA. (c/g) and (d/h): Symmetrized and antisymmetrized images of (a/e) and (b/f). The contrast of (c/g) is much larger than (d/h), same as $\mathrm{TaS_2}$.}
\end{figure*}

In this study, we have used single crystal samples of $\mathrm{TaS_2}$, $\mathrm{NbSe_2}$ and $\mathrm{TaSe_2}$ prepared by iodine vapor transport method.
The STM experiments were performed using our home-built systems, including a variable-temperature STM (6 to 300~K) and a low-temperature STM (5.3~K).
In both systems, the samples were cleaved at room temperature in ultra-high vacuum before being loaded into the STM. Etched tungsten tips, prepared \emph{in situ} by field emission against gold surfaces, are used to obtain high quality tunneling junctions together with spatial resolution at atomic scale.
We have examined three 2H-TMDs, $\mathrm{TaS_2}$, $\mathrm{TaSe_2}$, and $\mathrm{NbSe_2}$.
Fig.~\ref{fig1}a is a representative constant current topographic image of a 2H-$\mathrm{TaS_2}$ surface showing both the atomic corrugation and the 3 $\times$ 3 CDW superlattice. Fig.~\ref{fig1}b shows crystal structure with the neutral cleave plane (dashed line) between the two layers of chalcogen atoms.
The atomic lattice shown in Fig.~\ref{fig1}a is the resulting triangular chalcogen lattice, while the evident 3 $\times$ 3 superlattice is due to the formation of the CDW. The Fourier transform (Fig.~\ref{fig1}c) of Fig.~\ref{fig1}a shows that the atomic and the CDW signals are the main features of this surface.
After excluding the atomic corrugation by Fourier filtering, we find that the height of the top layer of atoms vary on the order of picometer in the z-direction due to the formation of CDW, for all three of the 2H-TMDs.
This change is consistent with the distortion measured by both x-ray diffraction \cite{Malliakas2013} and neutron scattering experiments \cite{Moncton1977} strongly indicating that our surface sensitive measurements of the CDW shows similar behavior to that seen in the bulk.
The strong commensurability between the CDW and the lattice in the STM images agrees well with that measured by scattering experiments.
This is in agreement with the strong $\boldsymbol{q}$-dependent el-ph interaction being involved in the CDW of these materials \cite{Weber2011}.

In order to probe the differences between the filled and empty states of the CDW phases in these materials topographic images were taken at opposite sample bias voltages.
In Figs.~\ref{fig2} and \ref{fig3}, our topographic images obtained on 2H-$\mathrm{TaS_2}$, $\mathrm{NbSe_2}$ and $\mathrm{TaSe_2}$ are shown.
Figs.~\ref{fig2}a and \ref{fig2}b are images of the filled and empty states (as indicated by negative and positive sample biases) in $\mathrm{TaS_2}$ while Figs.~\ref{fig3}a-b and \ref{fig3}e-f are images of the empty and filled states in $\mathrm{NbSe_2}$ and $\mathrm{TaSe_2}$.
Prior to analysis, these images are processed with a drift-correction algorithm \cite{Lawler2010} so that images taken at different times can be aligned on a subatomic length scale. Surface defects are used to ensure accuracy of the drift correction algorithm.
For images of TaS$_2$, line subtraction is applied to remove a low frequency vibrational noise.
This process, combined with the high thermal stability of our experiment, allows for an extremely precise measurement of position (with less than 50 pm of drift per day) and enables us to uncover the true structure of the CDW state in these materials for the first time.

Comparing our results with those from known CDW systems \cite{Mallet1999,Brun2009,Rahnejat2011}, which include STM results on $\mathrm{NbSe_2}$ \cite{Mallet1996}, we find qualitative discrepancies.
In our measurements, the filled/empty states topographic images of the TMD samples do not show the expected out-of-phase or contrast inversion that the Peierls model predicts.
This contrasts with previous observations of CDW systems under STM imaging.
Instead, the topographic images taken at +50 and -50~mV sample bias voltages on $\mathrm{TaS_2}$ are very similar to each other when one compares them with the naked eye.
In order to rigorously check this similarity, we have calculated the normalized cross correlation coefficients (NCC) between the two images, $NCC = \left.\langle(a_{ij}-\bar{a}) (b_{ij}-\bar{b})\rangle\middle/(\sigma_a \sigma_b)\right.$, where $\mu$ and $\sigma$ are the mean and the standard deviation of the two images ($a_{ij}$ and $b_{ij}$).
NCC is expected to be +1 for two identical images, and $-$1 for two images with perfect contrast inversion.
For an eMod-dominating CDW, where spatial inversion between filled and empty states is expected, one would expect the NCC between the filled and empty states images to be close to $-$1.
What we find when we carry out this analysis on $\mathrm{TaS_2}$, $\mathrm{NbSe_2}$ and $\mathrm{TaSe_2}$ is that the NCCs are +0.62, +0.79 and +0.95 respectively, showing that the majority of the topographic images are in-phase between the two biases.
Considering that only the CDW corrugation will be out-of-phase while the atomic corrugation is actually in-phase, we can remove the atomic signals by Fourier filtering them and keep only the CDW signal. After removing the atomic signals, the measured NCC are still +0.37, +0.67 and +0.97 for $\mathrm{TaS_2}$, $\mathrm{NbSe_2}$ and $\mathrm{TaSe_2}$ respectively, in sharp contrast to the $-$1 expected for a traditional CDW.
It is important to note that this result differs from previous STM results on $\mathrm{NbSe_2}$ \cite{Mallet1996} where spatially out-of-phase electron and hole components were seen.
This discrepancy is due to our superior ability to control and correct for drift at the subatomic level.

Interestingly our measured positive valued NCCs do not agree with the conventional pure electronic CDW model \cite{Gruner1994} and instead show that there is another non-electronic, lattice effect that is important to the formation of the 2D CDW in TMDs.
PLD is a natural explanation for the positive NCCs since STM measurement is a combination of the atomic structure (via the tip-sample distance) and the integrated density of states.
If the ionic cores, around which the valence electron clouds are centered, are displaced with the CDW transition, i.e. lattice distortion occurs, then the filled-state and empty-state topographic images will be in-phase with each other.
This is because they only enter into the tunneling current through the electron-hole symmetric matrix element.
In other words, if there is only a periodic lattice distortion then the filled-state and empty-state topographic images will be the same, except with a possible difference in the amplitudes of  the height profiles, which will not affect the NCC.
The expected behavior of a PLD dominates what we see in our STM images of the 2H-TMDs.

By calculating the symmetrized (Figs.~\ref{fig2}c, \ref{fig3}c and \ref{fig3}g, $S=(a+b)/2$ where $a$ and $b$ are the images of filled and empty states) and anti-symmetrized (Figs.~\ref{fig2}d, \ref{fig3}d and \ref{fig3}h, $AS=(a-b)/2$) images, we can directly access the PLD and eMod parts of the data respectively \footnote{See Supplemental Material for detailed information}.
Results of these calculations, shown in Figs.~\ref{fig2} and \ref{fig3} using the same height scaling, clearly show that the symmetrized channel has a larger z-range than that of the anti-symmetrized one, agreeing with our measured NCCs above.
The two channels ($S$ and $AS$) reveal the decomposed, spatially resolved, PLD and eMod components of the 3 $\times$ 3 superlattice for the first time. We find similar signals for the two other members of the 2H family that we have measured, $\mathrm{NbSe_2}$ and $\mathrm{TaSe_2}$.
In the framework of Peierls instability, the lattice distortion acts as a secondary effect which has to be minimized because it increases the elastic energy of the ground state.
Contrasting with the results of the previous STM study \cite{Mallet1996}, we have demonstrated that the PLD is playing an important, if not the primary, role in the CDW formation in 2H-TMDs. This is calling out a strong entanglement of the PLD and the eMod, which is not to be expected in a purely electronic Peierls picture.

Our results agree with the calculations which show that without the lattice involvement a CDW in these systems cannot exist. It has long been argued that without lattice distortions electron correlations are too strong for a purely electronic CDW to form \cite{Chan1973} and that there must be corresponding lattice distortions.
Our observations clearly show that these lattice distortions are present and dominate the observed structure in the 2H-TMDs, which is in contrast to previous reports to the contrary \cite{Mallet1996}.
In addition, recent theoretical calculations in this family of materials have shown that the observed CDW wave vectors do not correspond to the peaks in the electronic susceptibility \cite{Johannes2006}.
This calls for a more complicated model that includes nontrivial phonon involvement.
Such complicated origins of the charge ordering transition in such a simple family of materials demonstrates the necessity of taking the lattice contribution into account in describing the phase diagrams of everything from the simple systems like TMDs, to the more complex correlated electron materials like cuprates and manganites.

The symmetrization process rejects the antisymmetric eMod signal \footnote{See Fig.~S2 in Supplemental Material}, while keeping the PLD contribution since it enters the images of both the filled and empty states.
It is possible, however, that the anti-symmetrized images do contain more information than just that of the eMod.
In this sense, more quantitative modeling, such as first principle calculation, would be necessary to compare the relative intensities of PLD and eMod we present here.
As we mentioned earlier, the topography is only sensitive to the energy window which is set by tunneling bias voltage ($\pm$25 to 200 mV in our experiments), so higher energy studies might reveal a more complete picture of CDW in 2H-TMD.
Nonetheless, our study provides strong evidence that the 2D CDWs in the trigonal prismatic dichalcogenides go beyond the conventional Peierls picture.

In conclusion, we have demonstrated the importance of a high resolution, high stability, low drift STM to the understanding of the structure and origins of CDWs in the TMDs.
We have shown that atomic length scale measurements are necessary in order to resolve and separate the structures of both the PLD and eMod in $\mathrm{TaS_2}$, $\mathrm{NbSe_2}$ and $\mathrm{TaSe_2}$.
This shows that the PLD dominates the CDW transition for the entire family of 2H-TMD and that the eMod, which has been previously been seen as the dominating mechanism, plays a secondary role instead.
Our results indicate that weak coupling electron driven CDW is unlikely to exist in the 2H-TMDs and that the PLD plays an important role, which is in excellent qualitative agreement with the recent theoretical studies \cite{Johannes2008} and provides quantitative benchmarks to refine our understanding.

\begin{acknowledgments}
The authors thank Dmitry~Reznik, Dan~Dessau, Yue~Cao and Ray~Osborn for useful discussions. K.M. acknowledges the support by A.~P.~Sloan foundation. Y.S. and W.L. acknowledge the support by the National Key Basic Research (Grant No.~2011CBA00111) and the Joint Funds of the National Natural Science Foundation of China and the Chinese Academy of Sciences' Large-Scale Scientific Facility (Grant No.~U1232139).
\end{acknowledgments}


%

\end{document}


\title{Supplementary material for: \\Microscopic evidences for strong periodic lattice distortion in two-dimensional charge-density wave systems}

\author{Jixia Dai}
\email[]{daij@colorado.edu}
\affiliation{Department of Physics, University of Colorado at Boulder, Boulder, CO 80309, USA}

\author{Eduardo Calleja}
\affiliation{Department of Physics, University of Colorado at Boulder, Boulder, CO 80309, USA}

\author{Jacob Alldredge}
\affiliation{Department of Physics, University of Colorado at Boulder, Boulder, CO 80309, USA}

\author{Xiangde Zhu}
\affiliation{High Magnetic Field Laboratory, CAS, Hefei 230031, China}

\author{Lijun Li}
\affiliation{Key Laboratory of Materials Physics, Institute of Solid State Physics, CAS, Hefei 230031, China}

\author{Wenjian Lu}
\affiliation{Key Laboratory of Materials Physics, Institute of Solid State Physics, CAS, Hefei 230031, China}

\author{Yuping Sun}
\affiliation{Key Laboratory of Materials Physics, Institute of Solid State Physics, CAS, Hefei 230031, China}
\affiliation{High Magnetic Field Laboratory, CAS, Hefei 230031, China}

\author{Thomas Wolf}
\affiliation{Karlsruher Institut f\"{u}r Technologie, Institut f\"{u}r Festk\"{o}rperphysik, P.O.B. 3640, D-76021 Karlsruhe, Germany}

\author{Helmuth Berger}
\affiliation{Institute of Physics of Complex Matter, Ecole Polytechnique Federal de Lausanne, CH-1015 Lausanne, Switzerland}

\author{Kyle McElroy}
\email[]{Kyle.McElroy@colorado.edu}
\affiliation{Department of Physics, University of Colorado at Boulder, Boulder, CO 80309, USA}

\maketitle

\section{Local density of states calculation}
From the BCS-like mean field theory calculation (chapter 3 of Gr\"{u}ner's book, Ref.~\onlinecite{Gruner1994}), the ground state wave function is,
\begin{equation}\label{eqs:BCS_type_wave_function}
    \left|\phi_0\right\rangle = \left(\prod_{|k|<k_F}\gamma^\dag_{1,k}\gamma^\dag_{2,k}\right)\left|0\right\rangle
\end{equation}
where
\begin{equation}\label{eqs:Bogoliubov_transformation}
    \gamma_{1,k}=U_k e^{-i\phi/2}a_{1,k} - V_k e^{i\phi/2}a_{2,k}, \quad \gamma_{2,k}=V_k e^{-i\phi/2}a_{1,k} + V_k e^{i\phi/2}a_{2,k},
\end{equation}
and
\begin{equation}\label{eqs:Bogoliubov_transformation_solns1}
    V_k^2 = \frac{1}{2}\left(1+\frac{\epsilon_k}{E_k}\right), \quad U_k^2 = \frac{1}{2} \left(1 - \frac{\epsilon_k}{E_k}\right).
\end{equation}
The energy dispersion is,
\begin{equation}\label{eqs:BCS_type_energy_band}
    E_k=sign(k-k_F)\left(\epsilon_k^2 + \Delta^2\right),\quad \epsilon_k = \hbar v_F (k-k_F),
\end{equation}
From ground state wave function, one can derive the density of states with a single momentum state,
\begin{equation}\label{eqs:MF_DOS}
    \rho(x,k) = 1 - 2U_k V_k \cos(2k_F x +\phi),
\end{equation}
The equations above can be found in Gr\"{u}ner's book \cite{Gruner1994}.

\begin{figure}[htbp]
  \renewcommand\thefigure{S1}
  \includegraphics[width=3.25in]{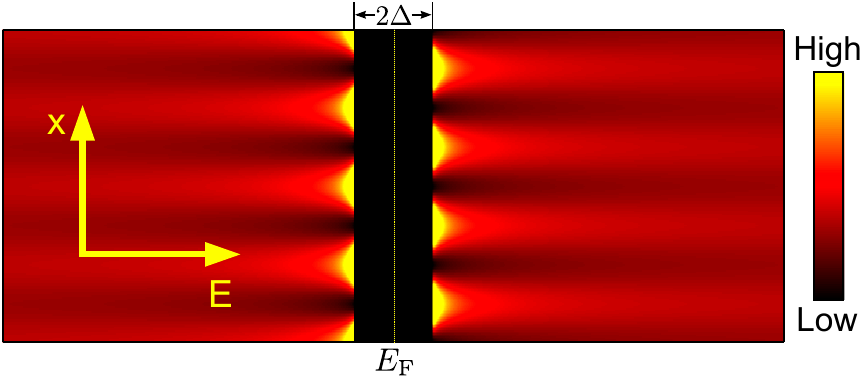}\\
  \caption{\label{figs:figS1}LDOS evolution with the periodicity of CDW. Three clear features are: 1. There is a gap around E$_\mathrm{F}$; 2. States around $E_\mathrm{F}$ are affected the most; 3. The LDOS enhancement alternates between the filled and empty sides.}
\end{figure}

\begin{figure}[htbp]
  \renewcommand\thefigure{S2}
  \includegraphics[width=3.25in]{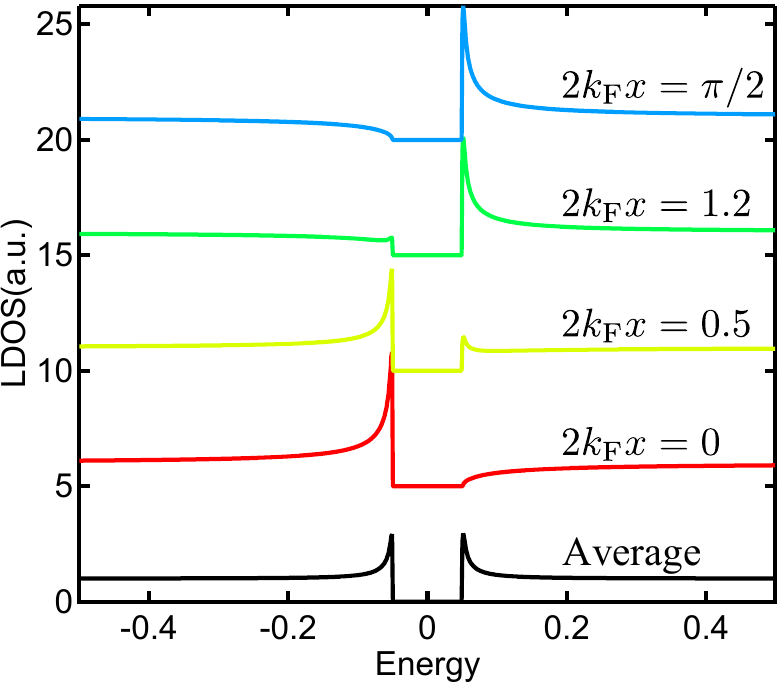}\\
  \caption{\label{figs:figS2}Representative LDOS with different phases of the CDW. The average LDOS is the same as the DOS predicted by BCS theory. The curves are offset for clarity.}
\end{figure}

To obtain local density of states (LDOS) which is what we measure with differential conductance (dI/dV), we need to convert momentum to energy,
\begin{equation}\label{eqs:MF_LDOS}
    \rho(x,E)= \rho(x, k(E))\frac{dk}{dE} =\left[1-\frac{|\Delta|}{E}\cos(2k_\mathrm{F}x+\phi)\right]\frac{|E|}{\sqrt{|E|^2-|\Delta|^2}},
\end{equation}
The second part of this LDOS is exactly the same as the DOS predicted by BCS theory, while the part in the bracket is an energy dependent spatial modulation term. This term implies that LDOS change brought by CDW transition concentrates around E$_\mathrm{F}$. To visualize this LDOS, we have plotted a simulated line-cut (Fig.~\ref{figs:figS1}) and several representative LDOS's with different phases (Fig.~\ref{figs:figS2}),

In the above simulation, the parameters we used are: $\hbar v_\mathrm{F}\!=\!1$, $k_\mathrm{F}\!=\!1$ and hence $\epsilon_\mathrm{F}\!=\!1$. We also used that $\phi\!=\!0$ and $\Delta\!=\!0.05$.
In order to avoid divide-by-zero issues and to include a physically meaningful temperature broadening, we have included a small imaginary part for the energy, $E\!=\!E_0+i\delta$.
For the calculation here we used $\delta\!=\!0.001$, which is a fair assumption for experiments at T$\ge$4K.

\section{Theory for symmetrization of topographic images}
The complete form of the tunneling current is as follows \cite{Chen1993},
\begin{equation}\label{eqs:tunnel_current}
    I = \frac{4\pi e}{\hbar}\int_{-\infty}^{+\infty} \left[f(E_\mathrm{F}-eV+\epsilon)-
    f(E_\mathrm{F}+\epsilon)\right] LDOS(E_\mathrm{F}-eV+\epsilon)
    \rho_\mathrm{t}(E_\mathrm{F}+\epsilon)|M|^2d\epsilon,
\end{equation}
where $f(E)$ is the Fermi-Dirac distribution function, $LDOS$ and $\rho_\mathrm{t}$ are density of states of sample and the tip, and $M$ is the tunneling matrix element connecting the tip and sample wave functions.
By assuming the energy window of interest is small compared to the other energy scales in the tunneling junction such as the work function $\phi$, the matrix element $M$ (following Bardeen, Ref.~\onlinecite{Bardeen1961}) is $M\sim e^{-\kappa s}$. Here $s$ is the tip-sample separation.
Furthermore, we can assume T = 0 and $\rho_\mathrm{t}\sim$ const. which gives us
\begin{equation}\label{eqs:sim_tunnel_current}
    I \propto e^{-2\kappa s}\int_0^{eV} LDOS(E_\mathrm{F} - \epsilon)d\epsilon
\end{equation}
Thus, we see that tunneling current in a STM junction is proportional to $e^{-2\kappa s}$ and the integrated density of states.

During the constant current imaging, the tip sweeps across a sample surface with the feedback active.
The height of the tip $z(x)$ closely follows the height of the surface $h(x)$ (defined as the position of the ionic cores and $s(x)=z(x)-h(x)$). By inverting the above equation, we get
\begin{equation}\label{eqs:topo}
    z(x)=h(x)+z_0 \ln I_0/I_s + z_0 \ln\int_0^{eV} LDOS(x, \epsilon)d\epsilon
\end{equation}

From the above equation, we see that tip height should follow $h(x)$ closely. More assumption with the integrated density of states can further reduce the above equation,
\begin{equation}\label{eqs:cdos}
    CDOS(x,V)=\int_0^{eV} LDOS(x, \epsilon)d\epsilon = C_0 (V)+C_1 (V)\cos\left(q_\mathrm{CDW} x\right)
\end{equation}
where $C_0$ is a spatial constant term and $C_1$ is a CDW related term. Corrugation terms ($C_1$) are usually small compared to $C_0$, so we can linearize the topographic density of states terms,
\begin{equation}\label{eqs:z_linearize}
    z(x,V)=h(x)+z_0 \ln I_0/I_s + z_0 \ln  C_0(V) + z_\mathrm{CDW}(V)\cos(q_\mathrm{CDW}x)
\end{equation}
where $z_\mathrm{CDW}$ is the amplitude of corrugation caused by CDW. This is true because any atomic scale variation, as verified experimentally, does not change with a change in the bias voltage so it can be absorbed in $h(x)$. Except for the atomic lattice component, we expect the topography to only include information relating to the CDW periodicity. This is supported by the data we have obtained.

In a pure electron modulation CDW, where the atomic locations are kept unchanged, the term $z_\mathrm{CDW}(V)$ will be anti-symmetric with respect to $V$, $z_\mathrm{CDW}(-V)=-z_\mathrm{CDW}(V)\equiv -z_\mathrm{A}(V)$. This is expected from our LDOS calculation in part I. However, if there is only a lattice distortion with no change in the electron-hole asymmetry, this term will be symmetric, $z_\mathrm{CDW}(-V)=z_\mathrm{CDW}(V) \equiv z_\mathrm{S}(V)$. Therefore, it is reasonable to include both components in a general case,
$z_\mathrm{CDW}(V)=z_\mathrm{A}(V)+z_\mathrm{S}(V)$, with $z_\mathrm{CDW}(-V)=-z_\mathrm{A}(V) + z_\mathrm{S}(V)$
where $z_\mathrm{A}$ and $z_\mathrm{S}$ are the contributions from electron modulation and periodic lattice distortion respectively. It can be found that,
\begin{equation}\label{eqs:zsa}
    z_\mathrm{S}(V) = \frac{z(x,V)+z(x,-V)}{2},\quad\text{and}\quad z_\mathrm{A}(V)=\frac{z(x,V)-z(x,-V)}{2}
\end{equation}
Thus, we have theoretically demonstrated how it is possible to image the electron modulation and lattice distortion separately by calculating these two terms.
Of course if the CDW components were large compared to the atomic position scale or if the exponential decay of the states were to vary spatially these approximations would be less valid. We have verified that both of these approximations are valid in the systems studied here.

%